\documentstyle[aps, epsfig]{revtex}
\tighten
\draft
\newcommand{\ket}[1]{\left | \, #1 \right \rangle}
\title{FOUR
PHOTON ENTANGLEMENT
 FROM DOWN CONVERSION }
\author{ Harald Weinfurter$^1$ and Marek \.Zukowski$^2$}

  \address{$^1$ Sektion Physik,
Ludwig-Maximilians-Universit\"at, D-80797 M\"{u}nchen, Germany,
 and \\ Max-Planck-Insitut f\"{u}r Quantenoptik, D-85748 Garching, Germany}
 \address{$^2$Instytut Fizyki Teoretycznej i Astrofizyki Uniwersytet
  Gda\'nski, PL-80-952 Gda\'nsk, Poland} \date{submitted Feb. 9, 2001}

\begin{document}
\maketitle

\begin{abstract}
 Double-pair emission from type-II parametric down conversion
results in a highly entangled 4-photon state. Due to
interference, which is similar to bunching from thermal emission,
this state is not simply a product of two pairs. The observation
of this state can be achieved by splitting the two emission modes
at beam splitters and subsequent detection of a photon in each
output. Here we describe the features of this state and give
a Bell theorem for a 4-photon test of local realistic
hidden variable theories.
\end{abstract}

%\twocolumn


%\section{Introduction}
Parametric down conversion has proven to be the best source of
entangled photon pairs so far in an ever increasing number of
experiments on the foundations of quantum mechanics
\cite{expfoundation} and in the new field of quantum
communication. Experimental realizations of concepts like
entanglement based quantum cryptography \cite{expentbasedQC},
quantum teleportation \cite{expQT} and its variations \cite{varQT}
demonstrated the usability of this source. New proposals for
quantum communication schemes \cite{multipartyQC} and, of course,
for improved tests of local hidden variable theories initiated the
quest for entangled multi-photon states. Interference of photons
generated by independent down conversion processes enabled the
first demonstration of a three-photon Greenberger-Horne-Zeilinger
(GHZ)-argument \cite{expGHZ} and, quite recently, even the
observation of a four-photon GHZ-state\cite{exp4GHZ}.

In this report we show that four photon entanglement can be
obtained directly from type-II parametric down conversion. Instead
of sophisticated but fragile interferometric set-ups, we utilize
bosonic interference in a double-pair emission process. This
effect causes strong correlations between measurement results of
the 4 photons and renders type-II down conversion a valuable tool
for new multi party quantum communication schemes. The analysis of
the entanglement inherent in the four photon emission leads us to
a new form of inequality distinguishing local hidden variable
theories from quantum mechanics, and demonstrates its potentiality
for experiments on the foundations of quantum mechanics.

%\section{the process}

In type-II parametric down conversion \cite{type-II} multiple
emission events during a single pump pulse lead to the following
state
\begin{equation}
Z \exp{\left(-i\alpha(a^*_Vb^*_H+a^*_Hb^*_V)\right)}\ket{0},
\label{1}
\end{equation}
where $Z$ is a normalization constant, $\alpha$ is proportional to
the pulse amplitude, and where $a^*_V$ is the creation operator of
a photon with vertical polarization in mode $a$, etc. (Fig. 1). We
are interested only in 4 photon effects, i.e. the emission of two
pairs. Then only the term in (\ref{1}) proportional to
\begin{equation}
(a^*_Hb^*_V+a^*_Vb^*_H)^2\ket{0}, \label{2}
\end{equation}
is relevant.
The particle interpretation of this term can be obtained by its
expansion
\begin{equation}
(a^{*2}_Hb^{*2}_V+a^{*2}_Vb^{*2}_H + 2a^*_V a^*_H b^*_V
b^*_H)\ket{0}, \label{3}
\end{equation}
and is given by the following superposition  of photon number
states
\begin{equation}
\ket{2H_a, 2V_b}+\ket{2V_a,2H_b}+\ket{1H_a,1V_a,1H_b,1V_b},
\label{4}
\end{equation}
where e.g. $2H_a$ means 2 H polarized photons in the beam $a$.

One should stress here that this type of description is valid only
for down conversion emissions, which is detected behind filters
endowed with a frequency band, which is narrower than that of the
pumping fields \cite{ZZW}. If a wide band down-conversion is
accepted then such a state is effective only if counts at the
detectors are treated as coincidences, when they occur within time
windows narrower than the inverse of the bandwidth of the
radiation \cite{ZZHE}. If such conditions are not met, then the
four photon events are essentially emissions of two independent,
entangled pairs, with the entanglement existing only within each
pair.

Let us pass the four photon state via two polarization independent
$50-50 $ beam splitters. For simplicity we assume that at the beam
splitters $a$ is transformed into ${1\over\sqrt{2}}{(a+a')}$ and
$b$ into ${1\over\sqrt{2}}{(b+b')}$, with prime denoting the
reflected beam. One can expand the expression (\ref{4}), and then
extract only those terms that lead to 4 photon coincidence behind
the two beam splitters, i.e. only those terms for which there is
one photon in each of the beams. The resulting component of the
full state is given by
\begin{equation}
\left[4\left(a^*_Ha'^*_Hb^*_Vb'^*_V+a^*_Va'^*_Vb^*_Hb'^*_H\right)
+2(a^*_Ha'^*_V+a^*_Va'^*_H)(b^*_Vb'^*_H+b^*_Hb'^*_V) \right]\ket{0}. \label{5}
\end{equation}
The first term represents a 4-photon GHZ state, whereas the second
one is a product state of two EPR-Bohm states (the $\ket{\Psi^+}$
Bell states in polarizations $H$ and $V$). This, after the
normalization, can be symbolically written as
\begin{equation} \sqrt{2/3}
\ket{GHZ}_{aa'bb'}+\sqrt{1/3}\ket{EPR}_{aa'}\ket{EPR}_{bb'}. \label{6}
\end{equation}
For additional simplicity of the presentation we also rotate the
polarizations in the beams $a$ and $a'$ by $90^\circ$. Thus now
our initial state is given by (\ref{6}) with the GHZ state in its
standard form, resulting in the state
\begin{eqnarray}&
\sqrt{1/3} \left(  \left. \ket{VVVV}_{aa'bb'}+\ket{HHHH}_{aa'bb'}
\right. \right.& \nonumber \\ &\left.
 +\frac{1}{2} \left( \ket{HVHV}_{aa'bb'}+\ket{HVVH}_{aa'bb'}+
 \ket{VHHV}_{aa'bb'}+\ket{VHVH}_{aa'bb'}
\right) \right) .& \label{7}
\end{eqnarray}

In order to demonstrate the entanglement of this state let us
analyze polarization correlation measurements involving all four
exit ports of the beam splitters, where the actual observables to
be measured are elliptic polarizations with main axis of the
polarization ellipse at $45^\circ$. Such observables are of
dichotomic nature, i.e. endowed with two valued spectrum
$k=+1,-1$, and are defined for each spatial propagation mode
$x=a,a',b,b'$ by their eigenstates
\begin{equation}
\sqrt{1/2}
\ket{V}_{x}+k e^{-i\phi_x}\sqrt{1/2}\ket{H}_{x}=\ket{k, \phi_x}. \label{8}
\end{equation}

The probability amplitudes for the results $k,l,m,n=\pm1$ at the
detector stations in the beams $a,a',b,b'$, under local phase
settings $\phi_a,\phi_{a'},\phi_{b},\phi_{b'}$, respectively, are
given by
\begin{equation}
\frac{1}{4\sqrt{3}}\left[1+klmn\,e^{i\sum\phi}+
\frac{1}{2}(k\,e^{i\phi_a}+l\,e^{\phi_{a'}})
(m\,e^{i\phi_b}+n\,e^{i\phi_{b'}})\right] , \label{9}
\end{equation}
where $\sum\phi$ stands for the sum of all local phases. Therefore
the probability to get  a
particular set of results $(k,l,m,n)$ is given by
\begin{eqnarray}&
P(k,l,m,n|\phi_a,\phi_{a'},\phi_{b},\phi_{b'})& \nonumber\\ &
=
\frac{1}{16}\left[\frac{2}{3}(1+klmn\cos{\sum\phi})\right.&
\nonumber \\ &\left.+\frac{1}{3} (1+kl\cos{(\phi_a-\phi_{a'})})
(1+mn\cos{(\phi_b-\phi_{b'})})\right.& \nonumber\\ &\left.
+\frac{1}{3}Re\left(
(1+klmn\,e^{i\sum\phi})(k\,e^{i\phi_a}+l\,e^{i\phi_{a'}})
(m\,e^{i\phi_b}+n\,e^{\phi_{b'}})\right)\right] .& \label{11}
\end{eqnarray}
The last term is written in the form of a real part of a complex
function to shorten the expression.

%As we shall see, surprisingly, the last term does not contribute
%to the correlation function for the four photon counts.

The correlation function is defined as the mean value of the
product of the four local results
\begin{eqnarray}&
E(\phi_a,\phi_{a'},\phi_{b},\phi_{b'})=
\sum_{k=\pm1}\sum_{l=\pm1}\sum_{m=\pm1}\sum_{n=\pm1}
klmn\,P(k,l,m,n|\phi_a,\phi_{a'},\phi_{b},\phi_{b'}) \;.&
 \label{12}
\end{eqnarray}
Its explicit form for the considered process is given by
\begin{eqnarray}&
E(\phi_a,\phi_{a'},\phi_{b},\phi_{b'})= \frac{2}{3}
\cos{\sum\phi}+\frac{1}{3}\cos{(\phi_a-\phi_{a'})}
\cos{(\phi_{b}-\phi_{b'})} \;.&  \label{13}
\end{eqnarray}
Only the first two terms of the probabilities (\ref{11})
contribute to the correlation function, and the function is itself
a weighted sum of the GHZ correlation function (the first term)
and a product of two EPR-Bell correlation functions. The last term
in (\ref{11}) gives a zero contribution to the correlation
function, because sums like $\sum_{klmn}ln$ vanish.

%\section{Bell theorem for all that}

The correlation function (\ref{13}) for the process has a more
complicated form than in the usual cases for GHZ-type states, but
the strong correlations for numerous phase settings clearly
indicate incompatibility with local realistic theories. When
inserted into Mermin-type Bell inequalities \cite{MERMIN} for four
particle systems, the violation is not too impressive, even for
optimal sets of local phases.
%When adapting the
%phases to this specific state one can obtain a somewhat stronger
%difference.
However, here we present a reasoning, involving Bell inequalities
of a new type \cite{ZUKOWSKI}, giving stronger inequalities for
distinguishing the validity of the different theories in a four
photon experiment.

In a local hidden variable theory a correlation function has to be
modeled by a construction of the following form (see e.g. \cite{BELL}, \cite{GHZ})
\begin{eqnarray}&
E_{LHV}(\phi_a,\phi_{a'},\phi_{b},\phi_{b'})=\int d\lambda \rho(\lambda)
I_a(\phi_a, \lambda)I_{a'}(\phi_{a'}, \lambda)
 I_b(\phi_b, \lambda)I_{b'}(\phi_{b'}, \lambda) \; ,
 \label{14b}
\end{eqnarray}
where $\lambda$ represents an arbitrary set of values of local
hidden variables, $\rho(\lambda)$ their probabilistic
distribution, and $I_x(\phi_x, \lambda)=\pm1$ $(x=a,a',b,b')$
represent the predetermined values of the measurements. Their
values depend on the set of hidden variables and on the value of
the {\it local} phase settings.

We start with allowing each observer of beam $x\;(=a,a',b,b')$ to
choose, just like in the standard cases of the Bell and GHZ
theorems \cite{BELL}, \cite{GHZ}, between two values $\phi_x^1$
and $\phi_x^2$ of the local phase settings.

The
formula for the LHV correlation function for the chosen settings is given by
\begin{eqnarray}&
E_{LHV}(\phi_a^p,\phi_{a'}^q,\phi_{b}^r,\phi_{b'}^s)
=\int d\lambda \rho(\lambda)
I_a(\phi_a^p, \lambda)I_{a'}(\phi_{a'}^q, \lambda)
 I_b(\phi_b^r, \lambda)I_{b'}(\phi_{b'}^s , \lambda)
 \label{14}
\end{eqnarray}
with $p,q,r,s=1,2$.
It is important to stress that one must consider arbitrary LHV
correlation functions. The only constraint being their structure
given by (\ref{14}).

One can treat the full set of the LHV predictions as a four index
tensor $\hat{E}_{LHV}$, with the indices $p,q,r,s=1,2$, built out
of the tensorial products of two dimensional real vectors
${\bf{v}}^{\lambda}_x=(I_{x}(\phi_{x}^1,\lambda),I_{x}(\phi_{x}^2,\lambda))$,
which represent the two possible results of a given observer
 for the given value of the hidden variable:
\begin{eqnarray}
& \hat{E}_{LHV}=\int d\lambda \rho(\lambda) {\bf{v}}^{\lambda}_a
\otimes {\bf{v}}^{\lambda}_{a'} \otimes {\bf{v}}^{\lambda}_b
\otimes {\bf{v}}^{\lambda}_{b'}
 \label{14a}
\end{eqnarray}

The actual values of the components of the two dimensional vectors
$(I_{x}(\phi_{x}^1,\lambda),I_{x}(\phi_{x}^2,\lambda))$ can be
equal to only either $(1,1)$, or $(1,-1)$, or $(-1,-1)$, or
finally $(-1,1)$. Let us denote these four possible vectors by
$\bf{v}^j_x$ with $j=1,2,3,4$, respectively. Thus, the LHV
correlation function (tensor) can be simplified to a discrete sum
over hidden probabilities $p_{k,l,m,n}$ of the tensorial products
of all possible measurement results.
\begin{eqnarray}
& \hat{E}_{LHV}=\sum_{k,l,m,n=1,...,4} \, p_{k,l,m,n}\,\bf{v}^k_a
\otimes\bf{v}^l_{a'} \otimes\bf{v}^m_b \otimes\bf{v}^n_{b'}
 \label{18}
\end{eqnarray}

%\vspace{1cm}

A further simplification of the tensor is possible since
$(-1,-1)=-(1,1)$ and $(-1,1)=-(1,-1)$, or in other words
$\bf{v}^{k+2}_x=-\bf{v}^{k}_x$.
The tensorial products $\bf{v}^k_a
\otimes\bf{v}^l_{a'} \otimes\bf{v}^m_b \otimes\bf{v}^n_{b'}$
with $k,l,m,n=1,2$ form
a complete orthogonal (product)
basis in the (real) Hilbert space of  tensors $
R^2\otimes R^2 \otimes R^2\otimes R^2$.
One can thus rewrite the expansion
(\ref{18}) so that it becomes an expansion in terms of the
aforementioned basis
\begin{eqnarray}
& \hat{E}_{LHV}=\sum_{k,l,m,n=1,2}\, c_{k,l,m,n}\,\bf{v}^k_a
\otimes\bf{v}^l_{a'} \otimes\bf{v}^m_b \otimes\bf{v}^n_{b'} \; .
 %& \nonumber\\
\label{19}
\end{eqnarray}
The relation between the coefficients in
(\ref{19}) and the probabilities of (\ref{18}) is given by
\begin{eqnarray}&
c_{k,l,m,n}=p_{k,l,m,n}-p_{k+2,l,m,n}-p_{k,l+2,m,n}-...& \nonumber
\\ &+p_{k+2,l+2,m,n}+...
-p_{k+2,l+2,m+2,n}+...+p_{k+2,l+2,m+2,m+2} \; .&
\end{eqnarray}
The expansion coefficients are of
course {\it unique}, and since  $\sum_{k,l,m,n=1,...,4}
p_{k,l,m,n}=1$ they satisfy the following inequality
\begin{eqnarray}
& \sum_{k,l,m,n=1,2} |c_{k,l,m,n}|\leq1 \; .
% & \nonumber\\
\label{20}
\end{eqnarray}
This inequality is a necessary condition for the local realistic
description to hold, and thus gives the handle for evaluating the
validity of this class of theories. It should be stressed that one
can also show that the inequality (\ref{20}) is also a sufficient
condition (see \cite{ZUKOWSKI}).

To compare the structure of the possible LHV correlation functions
with our quantum one (\ref{13}), let us, in order to simplify the
analysis, choose specific values for the local phase settings.
First, the observer of beam $a$ will be allowed the choice between
$\phi_a^1=0$ and $\phi_a^2=\pi/2$. The other observers
($y=a',b,b'$) can choose between $\phi_y^1=-\pi/4$ and
$\phi_y^2=\pi/4$. Next, one can expand the quantum function
(\ref{13}) into a sum of products of sine and cosine functions of
{\it single} phases
\begin{eqnarray}&
E(\phi_a,\phi_{a'},\phi_{b},\phi_{b'})& \nonumber \\ &=
c_ac_{a'}c_bc_{b'}
+s_as_{a'}s_bs_{b'}-\frac{1}{3}(s_as_{a'}c_bc_{b'}+c_ac_{a'}s_bs_{b'})&
\nonumber \\ &-\frac{2}{3}(s_ac_{a'}s_bc_{b'}+c_as_{a'}c_bs_{b'}
+c_as_{a'}s_bc_{b'}+s_ac_{a'}c_bs_{b'})& \; ,
 \label{21}
\end{eqnarray}
where $s_x=\sin{\phi_x}$ and $c_x=\cos{\phi_x}$.
%some changes below%
For each fixed set of four local settings one can calculate the
specific value of the quantum correlation function
$E(\phi_a^p,\phi_{a'}^q,\phi_{b}^r,\phi_{b'}^s).$
 We notice, that for
the specific phase settings given above one has ($y=a',b,b'$)
$$(\cos{(\phi_y^1)},\cos{(\phi_y^2)})= \frac{1}{\sqrt{2}}(1,1),$$
and
 $$(\sin{(\phi_y^1)},\sin{(\phi_y^2)})=
-\frac{1}{\sqrt{2}}(1,-1),$$ whereas
$$(\cos{(\phi_a^1)},\cos{(\phi_a^2)})=
(1,0)=\frac{1}{2}(1,1)+\frac{1}{2}(1,-1)$$ and
$$(\sin{(\phi_a^1)},\sin{(\phi_a^2)})=
(0,1)=\frac{1}{2}(1,1)-\frac{1}{2}(1,-1).$$ Therefore the quantum
predictions can be arranged to form a tensor, too, and it is easy
to write down its expansion in the product basis (the same basis
as in eq. (\ref{19}))
\begin{eqnarray}
& \hat{E}=\sum_{k,l,m,n=1,2} q_{k,l,m,n}\bf{v}^k_a
\otimes\bf{v}^l_{a'} \otimes\bf{v}^m_b \otimes\bf{v}^n_{b'} \; .
 %& \nonumber\\
\label{22}
\end{eqnarray}
The actual values of the expansion coefficients $q_{k,l,m,n}$ can
be straightforwardly obtained from (\ref{21}). However, note that
for the specific set of angles chosen above one has
\begin{eqnarray}
& \sum_{k,l,m,n=1,2} |q_{k,l,m,n}|=\frac{8}{3\sqrt{2}}>1 \; .
 %& \nonumber\\
\label{23}
\end{eqnarray}
Keeping in mind that the expansion in terms of basis vectors is
unique, the quantum correlation function out of which the tensor
$\hat{E}$ is built thus violates the necessary condition for local
realism (\ref{20}).

The quantum correlation function satisfies (\ref{20}) {\em only
if} it is multiplied by a scaling factor $v$ equal or smaller than
$\frac{3\sqrt{2}}{8}\approx53\%$, in other words, if one replaces
it by
\begin{equation}
E'(\phi_a,\phi_{a'},\phi_{b},\phi_{b'})=
 v E(\phi_a,\phi_{a'},\phi_{b},\phi_{b'}) \; .
\end{equation}
 In an interferometry experiment this
scaling parameter is directly related to the visibility
(contrast) of the interference pattern. Visibilities lower than one can be
interpreted as arising due to some noise contribution to the state.
%thus also gives the
%maximum noise contribution.
If one considers mixed states for the
system of the type
 \begin{equation}
\rho_v=(1-v)\rho_{noise}+v\ket{\bbox{\psi}}\langle\bbox{\psi}| \;,
 \end{equation}
where $\rho_{noise}=\frac{1}{16}\hat{I}$ represents completely
uncorrelated noise contribution, and $\ket{\bbox{\psi}}$ stands
for our pure state (\ref{6}), then the aforementioned critical $v$
gives the threshold beyond which no LHV model can resemble the
quantum predictions.

%\section*{Conclusion}

In conclusion, parametric type-II down conversion not only
produces entangled photon pairs, but also highly entangled four
photon states. The observation of these states is experimentally
much easier to achieve than for GHZ-type states. Here, the full
set of probabilities of possible LHV-predictions is compared with
the quantum predictions resulting in a significant distinction of
the theories.

The other interesting feature of the considered process is that
for a number of specific settings one obtains perfect four photon
correlations. E.g., for all local phases equal to zero the
correlation function is equal to $1$, whereas for
$\phi_b=\phi_{b'}$, $\phi_a-\phi_{a'}=\pi$ and
$\phi_{a'}+\phi_b=0$ it is equal to $-1$. This directly enables
one to transfer the standard protocols for entanglement based
quantum cryptography \cite{EkertBBM} to the four photon case
making multi party quantum key distribution and quantum
communication complexity schemes feasible.

MZ was supported by University of Gda\'nsk grant BW-5400-5-0032-0.
and the Erwin Schr\"odinger International Institute for
Mathematical Physics,  Austria. This work was supported by the
EU-Project QuComm (IST-1999-10033).

\end{document}